\documentclass[journal=jctcce,manuscript=article,layout=twocolumn]{achemso}
\setkeys{acs}{etalmode=truncate}
\setkeys{acs}{maxauthors=3}
\setkeys{acs}{articletitle=true}

\usepackage{amsmath}
\usepackage{bm}
\usepackage{braket}
\usepackage[capitalise]{cleveref}
\usepackage[inline]{enumitem}

\usepackage[colorinlistoftodos]{todonotes}

\newcommand{\HF}{\ensuremath{\textrm{HF}}}
\newcommand{\CI}{\ensuremath{\textrm{CI}}}
\newcommand{\CC}{\ensuremath{\textrm{CC}}}
\newcommand{\bohr}{\ensuremath{\mathrm{a_0}}}

\usepackage{bbold}
\usepackage{listings}
\usepackage{color}

\definecolor{dkgreen}{rgb}{0,0.6,0}
\definecolor{gray}{rgb}{0.5,0.5,0.5}
\definecolor{mauve}{rgb}{0.58,0,0.82}

\usepackage[symbol]{footmisc}

\author{James~S.~Spencer}
\affiliation{Dept.~of Physics, Imperial College London, South Kensington Campus, London SW7 2AZ, United Kingdom}
\alsoaffiliation{Dept.~of Materials, Imperial College London, South Kensington Campus, London SW7 2AZ, United Kingdom}
\author{Nick~S.~Blunt}
\affiliation{University Chemical Laboratory, Lensfield Road, Cambridge CB2 1EW, United Kingdom}
\alsoaffiliation{St John's College, St John's Street, Cambridge, CB2 1TP, United Kingdom}
\altaffiliation{Authors listed alphabetically}
\author{Seonghoon Choi}
\affiliation{University Chemical Laboratory, Lensfield Road, Cambridge CB2 1EW, United Kingdom}
\altaffiliation{Authors listed alphabetically}
\author{Ji\v{r}\'{\i} Etrych}
\affiliation{University Chemical Laboratory, Lensfield Road, Cambridge CB2 1EW, United Kingdom}
\altaffiliation{Authors listed alphabetically}
\author{Maria-Andreea Filip}
\affiliation{University Chemical Laboratory, Lensfield Road, Cambridge CB2 1EW, United Kingdom}
\altaffiliation{Authors listed alphabetically}
\author{W. M. C.~Foulkes}
\affiliation{Dept.~of Physics, Imperial College London, South Kensington Campus, London SW7 2AZ, United Kingdom}
\altaffiliation{Authors listed alphabetically}
\author{Ruth~S.T.~Franklin}
\affiliation{University Chemical Laboratory, Lensfield Road, Cambridge CB2 1EW, United Kingdom}
\altaffiliation{Authors listed alphabetically}
\author{Will~J.~Handley}
\affiliation{Astrophysics Group, Cavendish Laboratory, Cambridge, CB3 OHE, UK}
\alsoaffiliation{Kavli Institute for Cosmology, Madingley Road, Cambridge, CB3 0HA, UK}
\alsoaffiliation{Gonville \& Caius College, Trinity Street, Cambridge, CB2 1TA, UK}
\altaffiliation{Authors listed alphabetically}
\author{Fionn D.~Malone}
\affiliation{Dept.~of Physics, Imperial College London, South Kensington Campus, London SW7 2AZ, United Kingdom}
\alsoaffiliation{Quantum Simulations Group, Lawrence Livermore National Laboratory, Livermore, California 94550, USA}
\altaffiliation{Authors listed alphabetically}
\author{Verena A.~Neufeld}
\affiliation{University Chemical Laboratory, Lensfield Road, Cambridge CB2 1EW, United Kingdom}
\altaffiliation{Authors listed alphabetically}
\author{Roberto~Di Remigio}
\affiliation{Hylleraas Centre for Quantum Molecular Sciences, Department of Chemistry, University of Troms{\o} - The Arctic University of Norway, N-9037 Troms{\o}, Norway}
\alsoaffiliation{Department of Chemistry, Virginia Tech, Blacksburg, Virginia 24061, United States}
\altaffiliation{Authors listed alphabetically}
\author{Thomas W.~Rogers}
\affiliation{Dept.~of Physics, Imperial College London, South Kensington Campus, London SW7 2AZ, United Kingdom}
\altaffiliation{Authors listed alphabetically}
\author{Charles~J.C.~Scott}
\affiliation{University Chemical Laboratory, Lensfield Road, Cambridge CB2 1EW, United Kingdom}
\altaffiliation{Authors listed alphabetically}
\author{James~J.~Shepherd}
\affiliation{Chemistry Building, University of Iowa, Iowa, 52240, USA}
\altaffiliation{Authors listed alphabetically}
\author{William A.~Vigor}
\affiliation{Dept.~of Chemistry, Imperial College London, South Kensington Campus, London SW7 2AZ, United Kingdom}
\altaffiliation{Authors listed alphabetically}
\author{Joseph~Weston}
\affiliation{Dept.~of Physics, Imperial College London, South Kensington Campus, London SW7 2AZ, United Kingdom}
\altaffiliation{Authors listed alphabetically}
\author{RuQing Xu}
\affiliation{Department of Modern Physics, University of Science and Technology, Hefei, Anhui, 230026, China}
\altaffiliation{Authors listed alphabetically}
\author{Alex~J.W.~Thom}
\affiliation{University Chemical Laboratory, Lensfield Road, Cambridge CB2 1EW, United Kingdom}
\alsoaffiliation{Dept.~of Chemistry, Imperial College London, South Kensington Campus, London SW7 2AZ, United Kingdom}
\email{ajwt3@cam.ac.uk}

\title{The HANDE-QMC project: open-source stochastic quantum chemistry from the ground state up}

\begin{document}

\begin{abstract}
Building on the success of Quantum Monte Carlo techniques such as diffusion Monte Carlo, alternative stochastic approaches to solve
electronic structure problems have emerged over the last decade. The full configuration interaction quantum Monte Carlo (FCIQMC) method allows one to systematically approach the exact solution of such problems, for cases where very high accuracy is desired. The introduction of FCIQMC has subsequently led to the development of coupled cluster Monte Carlo (CCMC) and density matrix quantum Monte Carlo (DMQMC), allowing stochastic sampling of the coupled cluster wave function and the exact thermal density matrix, respectively. In this article we describe the HANDE-QMC code, an open-source implementation of FCIQMC, CCMC and DMQMC, including initiator and semi-stochastic adaptations. We describe our code and demonstrate its use on three example systems;
a molecule (nitric oxide), a model solid (the uniform electron gas), and a real solid (diamond). An illustrative tutorial is also included.
\end{abstract}
\section{Introduction}

Quantum Monte Carlo (QMC) methods, in their many forms, are among the most reliable and accurate tools available for the investigation of realistic quantum systems\cite{Foulkes2001}. QMC methods have existed for decades, including notable approaches such as variational Monte Carlo (VMC)\cite{mcmillan_ground_1965, Umrigar1988, Umrigar2007, Neuscamman2012, Neuscamman2016}, diffusion Monte Carlo (DMC)\cite{Grimm1971, Anderson1975, Umrigar1993, Foulkes2001, qmcpack} and auxiliary-field QMC (AFQMC)\cite{zhang_quantum_2003}; such methods typically have low scaling with system size, efficient large-scale parallelization, and systematic improvability, often allowing benchmark quality results in challenging systems.

A separate hierarchy exists in quantum chemistry, consisting of methods such as
coupled cluster (CC) theory\cite{cizek_correlation_1966}, M{\o}ller-Plesset
perturbation theory (MPPT),\cite{moller_note_1934} and configuration interaction (CI),
with full CI (FCI)\cite{knowles_new_1984} providing the exact benchmark within a
given single-particle basis set.
The scaling with the number of basis functions can be steep for these
methods: from $N^{4}$ for MP2 to exponential for FCI.
Various approaches to tackle the steep scaling wall have been proposed in the
literature: from adaptive selection algorithms\cite{Huron1973-wv,Scemama2013,
Schriber2016-xz, Tubman2016-rc, Holmes2016-qw, Garniron2017} and many-body expansions for
CI\cite{Eriksen2017-qg} to the exploitation of the locality of the one-electron basis set\cite{Saebo1993-qx}
for MP2 and CC.\cite{Hampel1996-yy, Riplinger2013-mz, Ziolkowski2010-oo}
Such approaches have been increasingly successful, now often allowing chemical
accuracy to be achieved for systems comprising thousands of basis functions.

In 2009, Booth, Thom and Alavi introduced the full configuration interaction quantum Monte Carlo (FCIQMC) method\cite{BoothAlavi_09JCP}.
The FCIQMC method allows essentially exact FCI results to be achieved for systems beyond the reach of traditional, exact FCI approaches;
in this respect, the method occupies a similar space to the density matrix renormalization group (DMRG) algorithm\cite{White1992,Chan2004,Amaya2015} and selected CI approaches.\cite{Huron1973-wv, Scemama2013, Schriber2016-xz, Tubman2016-rc, Holmes2016-qw, Garniron2017}
Employing a sparse and stochastic sampling of the FCI wave function greatly
reduces the memory requirements compared to exact approaches.
The introduction of FCIQMC has led to the development of several other related QMC methods, including coupled cluster Monte Carlo (CCMC)\cite{Thom_10PRL,SpencerThom_16JCP}, density matrix quantum Monte Carlo (DMQMC)\cite{blunt_density-matrix_2014,malone_interaction_2015}, model space quantum Monte Carlo (MSQMC)\cite{ten-no_stochastic_2013, Ohtsuka2015, Ten-no2017}, clock quantum Monte Carlo\cite{McClean2015-de}, driven-dissipative quantum Monte Carlo (DDQMC)\cite{Nagy2018-kh}, and several other variants, including multiple approaches for studying excited-state properties\cite{Booth2012_excited, ten-no_stochastic_2013, Humeniuk2014, Blunt2015}.

In this article we present HANDE-QMC (Highly Accurate N-DEterminant Quantum Monte Carlo), an open-source quantum chemistry code that performs several of the above quantum Monte Carlo methods. In particular, we have developed a highly-optimized and massively-parallelized package to perform state-of-the-art FCIQMC, CCMC and DMQMC simulations.

An overview of stochastic quantum chemistry methods in HANDE-QMC is given in \cref{sec:stochastic_qc}. \cref{sec:hande} describes the HANDE-QMC package, including implementation details, our development experiences, and analysis tools. Applications of FCIQMC, CCMC and DMQMC methods are contained in \cref{sec:results}. We conclude with a discussion in \cref{sec:discussion} with views on scientific software development and an outlook on future work. A tutorial on running HANDE is provided in the Supplementary Material.

\section{Stochastic quantum chemistry}
\label{sec:stochastic_qc}

\subsection{Full Configuration Interaction Quantum Monte Carlo}

The FCI ansatz for the ground state wavefunction is $\ket{\Psi_\CI} = \sum_{\textbf{i}}c_{\textbf{i}} \ket{D_{\textbf{i}}}$, where $\{D_{\textbf{i}}\}$ is the set of Slater determinants.
Noting that $(1-\delta\tau \hat{H})^N \ket{\Psi_0} \propto \ket{\Psi_\CI}$ as $N\rightarrow\infty$, where $\Psi_0$ is some arbitrary initial vector with $\braket{\Psi_0|\Psi_\CI} \ne 0$ and $\delta\tau$ is sufficiently small\cite{Spencer2012},
the coefficients $\{c_{\textbf{i}}\}$ can be found via an iterative process derived from a first-order solution to the imaginary-time Schr\"odinger equation\cite{BoothAlavi_09JCP}:
\begin{equation}
    c_{\textbf{i}}(\tau+\delta\tau) = c_{\textbf{i}}(\tau) - \delta\tau\sum_{\textbf{j}} \braket{D_{\textbf{i}}|\hat{H}|D_{\textbf{j}}} c_{\textbf{j}}(\tau).
    \label{eqn:fciqmc}
\end{equation}

A key insight is that the action of the Hamiltonian can be applied stochastically rather than deterministically: the wavefunction is discretized by using a set of particles with weight $\pm 1$ to represent the coefficients, and is evolved in imaginary time by stochastically creating new particles according to the Hamiltonian matrix (\cref{sec:sd_qmc}).
By starting with just particles on the Hartree--Fock determinant or a small number of determinants, the sparsity of the FCI wavefunction emerges naturally. The FCIQMC algorithm hence has substantially reduced memory requirements\cite{BoothAlavi_09JCP}
and is naturally scalable\cite{Booth2014}
in contrast to conventional Lanczos techniques.
The sign problem manifests itself in the competing in-phase and out-of-phase combinations of particles with positive and negative signs on the same determinant\cite{Spencer2012};
this is alleviated by exactly canceling particles of opposite sign on the same determinant, a process termed `annihilation'. This results in the distinctive population dynamics of an FCIQMC simulation, and a system-specific critical population is required to obtain a statistical representation of the correct FCI wavefunction\cite{Spencer2012}.
Once the ground-state FCI wavefunction has been reached, the population is controlled via a diagonal energy offset\cite{Umrigar1993,BoothAlavi_09JCP}
and statistics can be accumulated for the energy estimator and, if desired, other properties.

The stochastic efficiency of the algorithm (determined by the size of
statistical errors for a given computer time) can be improved by several approaches: using real weights, rather than integer weights, to represent particle amplitudes\cite{Petruzielo2012,Overy2014};
a semi-stochastic propagation, in which the action of the Hamiltonian in a small subspace of determinants is applied \emph{exactly}\cite{Petruzielo2012,Blunt2015_semistoch};
and more efficient sampling of the Hamiltonian by incorporating information about the magnitude of the Hamiltonian matrix elements into the selection probabilities\cite{holmes_efficient_2016,Neufeld2018}.

The initiator approximation\cite{ClelandAlavi_10JCP} (often referred to as i-FCIQMC) only permits new particles to be created on previously unoccupied determinants if the spawning determinant has a weight above a given threshold --- this introduces a systematic error which is reduced with increasing particle populations, but effectively reduces the severity of the sign problem. This simple modification has proven remarkably successful and permits FCI-quality calculations on Hilbert spaces orders of magnitude beyond exact FCI. 

\subsection{Coupled Cluster Monte Carlo}

The coupled cluster wavefunction ansatz is $\ket{\Psi_\CC} = N e^{\hat{T}} \ket{D_\HF}$, where $\hat{T}$ is the cluster operator containing all excitations up to a given truncation level, $N$ is a normalisation factor and $\ket{D_\HF}$ the Hartree--Fock determinant. For convenience, we rewrite the wavefunction ansatz as $\ket{\Psi_\CC} = t_\HF e^{\hat{T}/t_\HF} \ket{D_\HF}$, where $t_\HF$ is a weight on the Hartree--Fock determinant, and define $\hat{T} = \sum_{\textbf{i}}^\prime t_{\textbf{i}} \hat{a}_{\textbf{i}} $, where ${}^\prime$ restricts the sum to be up to the truncation level, $\hat{a}_{\textbf{i}}  $ is an excitation operator (excitor) such that $\hat{a}_{\textbf{i}} \ket{D_\HF}$ results in $\ket{D_{\textbf{i}}}$ and $t_{\textbf{i}}$ is the corresponding amplitude. Using the same first-order Euler approach as in FCIQMC gives a similar propagation equation:
\begin{equation}
    t_{\textbf{i}}(\tau+\delta\tau) = t_{\textbf{i}}(\tau) - \delta\tau \sum_{\textbf{j}} \braket{D_{\textbf{i}}|\hat{H}|D_{\textbf{j}}} \tilde{t}_{\textbf{j}}(\tau).
    \label{eqn:ccmc}
\end{equation}
The key difference between \cref{eqn:fciqmc,eqn:ccmc} is $\tilde{t}_{\textbf{j}} = \braket{D_{\textbf{j}} | \Psi_\CC}$ contains contributions from clusters of excitors\cite{Thom_10PRL}
whereas the FCI wavefunction is a simple linear combination.
This is tricky to evaluate efficiently and exactly each iteration. Instead, $\tilde{t}_{\textbf{j}}$ is sampled and individual contributions propagated separately\cite{Thom_10PRL,Spencer2018,Scott2017}.
Bar this complication, the coupled cluster wavefunction can be stochastically evolved using the same approach as used in FCIQMC.

\subsection{Density Matrix Quantum Monte Carlo}

FCIQMC and CCMC are both ground-state, zero-temperature methods (although excited-state variants of FCIQMC exist\cite{Booth2012_excited, ten-no_stochastic_2013, Humeniuk2014, Blunt2015}). The exact \emph{thermodynamic} properties of a quantum system in thermal equilibrium can be determined from the (unnormalized) $N$-particle density matrix, $\hat{\rho}(\beta) = e^{-\beta \hat{H}}$, where $\beta=1/k_B T$. A direct evaluation of $\hat{\rho}(\beta)$ requires knowledge of the full eigenspectrum of $\hat{H}$, a hopeless task for all but trivial systems. To make progress we note that the density matrix obeys the (symmetrized) Bloch equation
\begin{equation}
\frac{d\hat{\rho}}{d\beta} = -\frac{1}{2} \left[ \hat{H}\hat{\rho} + \hat{\rho}\hat{H} \right].
\label{eq:dmqmc}
\end{equation}
Representing $\hat{\rho}$ in the Slater determinant basis, $\rho_{\textbf{ij}} = \braket{D_{\textbf{i}} | \hat{\rho} | D_{\textbf{j}}}$ and again using a first-order update scheme results in similar update equations to FCIQMC and CCMC:
\begin{equation}
\begin{aligned}
\rho_{\textbf{ij}}(\beta+\delta\beta) &= \rho_{\textbf{ij}}(\beta) - \frac{\delta\beta}{2} \sum_{\textbf{k}} 
\left[
\braket{ D_{\textbf{i}} | \hat{H} | D_{\textbf{k}} } \rho_{\textbf{kj}}(\beta) \right. \\
 &+ \left. \rho_{\textbf{ik}}(\beta)\braket{ D_{\textbf{k}} | \hat{H} | D_{\textbf{j}} }
\right].
\end{aligned}
\end{equation}
It follows that elements of the density matrix can be updated stochastically in a similar fashion to FCIQMC and CCMC.
$\rho(\beta)$ is a single stochastic measure of the exact density matrix at inverse temperature $\beta$. Therefore, unlike FCIQMC and CCMC, multiple independent simulations must be performed in order to gather statistics at each temperature. The simplest starting point for a simulation is at $\beta=0$, where $\rho$ is the identity matrix. Each simulation (termed `$\beta$-loop') consists of sampling the identity matrix and propagating to the desired value of $\beta$. Averaging over multiple $\beta$-loops gives thermal properties at all temperatures in the range $[0,\beta]$.

While this scheme is exact (except for small and controllable errors due to finite $\delta \beta$), it suffers from the issue that important states at low temperature may not be sampled in the initial ($\beta=0$) density matrix, where all configurations are equally important\cite{malone_interaction_2015}. 
To overcome this, we write $\hat{H} = \hat{H}^0 + \hat{V}$ and define the auxiliary density matrix $\hat{f}(\tau) = e^{-(\beta-\tau)\hat{H}^0} \hat{\rho}(\tau)$ with the following properties:
\begin{gather}
\hat{f}(0) = e^{-\beta\hat{H}^0}, \\
\hat{f}(\beta) = \hat{\rho}(\beta), \\
\frac{d\hat{f}}{d\tau} = \hat{H}^0 \hat{f} - \hat{f} \hat{H}.
\label{eq:ipdmqmc}
\end{gather}
We see that with this form of density matrix we can begin the simulation from a mean-field solution defined by $\hat{H}_0$, which should (by construction) lead to a distribution containing the desired important states (such as the Hartree--Fock density matrix element) at low temperature. 
Furthermore, if $\hat{H}^0$ is a good mean field Hamiltonian then $e^{\beta\hat{H}^0} \hat{\rho}$ is a \emph{slowly varying} function of $\beta$, and is thus easier to sample.
Comparing \cref{eq:dmqmc,eq:ipdmqmc}, we see that $\hat{f}$ can be stochastically sampled in a similar fashion to DMQMC, with minor modifications relative to using the unsymmetrized Bloch equation\cite{blunt_density-matrix_2014}:
\begin{enumerate*}[label=\roman*)]
\item the choice of $\hat{H}^0$ changes the probability of killing a particle (\cref{sec:sd_qmc});
\item the $\tau=0$ initial configuration must be sampled according to $\hat{H}^0$ rather than the identity matrix;
\item evolving to $\tau=\beta$ gives a sample of the density matrix at inverse temperature $\beta$ \emph{only} - independent simulations must be performed to accumulate results at different temperatures.
\end{enumerate*}
We term this method interaction-picture DMQMC (IP-DMQMC).

\subsection{Commonality  between FCIQMC, CCMC and DMQMC}
\label{sec:sd_qmc}

FCIQMC, CCMC and DMQMC have more similarities than differences: the amplitudes within the wavefunction or density matrix are represented stochastically by a weight, or particle\bibnote{The original algorithm used integer weights. It was subsequently shown that floating-point weights greatly reduce the stochastic noise. HANDE-QMC uses fixed-precision for the weights such that both approaches can be straightforwardly handled.}. These stochastic amplitudes are sampled to produce states, which make up the wavefunction or density matrix. For FCIQMC (DMQMC), a state corresponds to a determinant (outer product of two determinants), and for CCMC corresponds to a term sampled from the cluster expansion corresponding to a single determinant.
The stochastic representation of the wavefunction or density matrix is evolved by
\begin{description}
\item[spawning] sampling the action of the Hamiltonian on each (occupied) state, which requires random selection of a state connected to the original state. The process of random selection (`excitation generation') is system-dependent, as it depends upon the connectivity of the Hamiltonian matrix;
    efficient sampling of the Hamiltonian has a substantial impact on the stochastic efficiency of a simulation\cite{Petruzielo2012,holmes_efficient_2016,Neufeld2018}. 
\item[death] killing each particle with probability proportional to its diagonal Hamiltonian matrix element.
\item[annihilation] combining particles on the same state and canceling out particles with the same absolute weight but opposite sign.
\end{description}
Energy estimators can be straightforwardly accumulated during the evolution process. 
A parallel implementation distributes states over multiple processors, each of which need only evolve its own set of states. The annihilation stage then requires an efficient process for determining to which processor a newly spawned particle should be sent\cite{Booth2014}. For CCMC an additional communication step is required to ensure that the sampling of products of amplitudes is unbiased\cite{Spencer2018}.

Hence, FCIQMC, CCMC and DMQMC share the majority of the core algorithms in the HANDE-QMC implementations.
The primary difference is the representation of the wavefunction or density matrix, and the action of the Hamiltonian in the representation. These differences reside in the outer-most loop of the algorithm and so do not hinder the re-use of components between the methods. This remains the case even for linked coupled cluster Monte Carlo, which applies the similarity-transformed Hamiltonian, $e^{-T}He^{T}$, and the interaction picture formulation of DMQMC.

It is important to note that this core paradigm also covers different approaches to propagation\cite{ten-no_stochastic_2013,Petruzielo2012,ClelandAlavi_10JCP,tubman_deterministic_2016},
the initiator approximation\cite{ClelandAlavi_10JCP,SpencerThom_16JCP,malone_accurate_2016},
excitation generators\cite{holmes_efficient_2016,Neufeld2018},
excited states and properties\cite{Overy2014,Blunt2015,Blunt2017},
and can naturally be applied to different wavefunction Ans\"atze\cite{shepherd_sen0_2016},
which can be added relatively straightforwardly on top of a core implementation of FCIQMC. Due to this, improvements in, say, excitation generators can be immediately used across all methods in HANDE.

\section{HANDE-QMC}
\label{sec:hande}

\subsection{Implementation}

HANDE-QMC is implemented in Fortran and takes advantage of the increased expressiveness provided by the Fortran 2003 and 2008 standards\bibnote{The use of Fortran 2003 and 2008 imposes a need for a recent Fortran compiler. Indeed, we have found bugs in both open-source and proprietary compilers and worked around them where possible.}. Parallelization over multiple processors is implemented using OpenMP (CCMC-only for intra-node shared memory communication) and MPI. Parallelization and the reusability of core procedures have been greatly aided by the use of pure procedures and minimal global state, especially for system and calculation data.

We attempt to use best-in-class libraries where possible. This allows for rapid
development and a focus on the core QMC algorithms. HANDE-QMC relies upon
MurmurHash2 for hashing operations\cite{smhasher}, dSFMT for high-quality pseudo-random number
generation\cite{dsfmt}, numerical libraries (cephes\cite{cephes}, LAPACK, ScaLAPACK,
TRLan\cite{trlan, Yamazaki2010-xq}) for special functions, matrix and vector procedures and Lanczos diagonalization, and HDF5 for file I/O\cite{hdf5}. The input file to HANDE-QMC is a Lua script\cite{Ierusalimschy2016-cz}; Lua is a lightweight scripting language designed for embedding in applications and can easily be used from Fortran codes via the AOTUS library\cite{aotus}. Some of the advantages of using a scripting language for the input file are detailed in \cref{sec:discussion}.

Calculation, system settings and other metadata are included in the output in the JSON format\cite{json}, providing a good compromise between human- and machine-readable output.

HANDE can be compiled either into a standalone binary or into a library, allowing it to be used directly from existing quantum chemistry packages. CMake\cite{cmake} is used for the build system, which allows for auto-detection of compilers, libraries and available settings in most cases. A legacy Makefile is also included for compiling HANDE in more complex environments where direct and fine-grained control over settings is useful.

Integrals for molecular and solid systems can be generated by Hartree--Fock calculations using standard quantum chemistry programs, such as Psi4\cite{parrish_psi4_2017}, HORTON\cite{HORTON}, PySCF\cite{Sun2018}, Q-Chem\cite{shao_advances_2015}, and MOLPRO\cite{MOLPRO}, in the plain-text FCIDUMP format. HANDE can convert the FCIDUMP file into an HDF5 file, which gives a substantial space saving and can be read in substantially more quickly. For example, an all-electron FCIDUMP for coronene in a Dunning cc-pVDZ basis\cite{Dunning_89JCP} is roughly 35GB in size and takes 1840.88 seconds to read into HANDE and initialise. When converted to HDF5 format, the resulting file is 3.6GB in size and initialising an identical calculation takes only 60.83 seconds. This is useful in maximizing resource utilization when performing large production-scale calculations on HPC facilities. The memory demands of the integrals are reduced by storing the two-electron integrals only once on each node using either the MPI-3 shared memory functionality or, for older MPI implementations, POSIX shared memory.

In common with several Monte Carlo methods, data points from consecutive iterations are not independent, as the population at a given iteration depends on the population at the previous iteration. This autocorrelation must be removed in order to obtain accurate estimates of the standard error arising from FCIQMC and CCMC simulations\bibnote{DMQMC averages over independent calculations and so does not suffer from a correlation issue.} and is most straightforwardly done via a reblocking analysis\cite{Flyvbjerg1989}. This can be performed as a post-processing step\cite{pyblock} but is also implemented as an on-the-fly algorithm\cite{kent_efficient_2007}, which enables calculations to be terminated once a desired statistical error has been reached.

It is often useful to continue an existing calculation; for example to accumulate more statistics to reduce the error bar, to save equilibration time when investigating the effect of calculation parameters or small geometry changes, or for debugging when the bug is only evident deep into a calculation.
To aid these use cases, calculations can be stored and resumed via the use of restart files. The state of the pseudo-random number generator is included in the restart files such that restarted calculations follow the same Markov chain as if they had been run in a single calculation assuming the same calculation setup is used. We use the HDF5 format and library for efficient I/O and compact file sizes. A key advantage of this approach is that it abstracts the data layout into a hierarchy (termed \emph{groups} and \emph{datasets}). This makes extending the restart file format to include additional information whilst maintaining backward compatibility with previous calculations particularly straightforward. Each calculation is labeled with a universally unique identifier (UUID)\cite{uuid}, stored in the restart file and included in the metadata of subsequent calculations. This is critical for tracing the provenance of data generated over multiple restarted calculations. 

Extensive user-level documentation is included in the HANDE-QMC package\cite{hande-doc} and details compilation, input options, running HANDE and calculation analysis. The documentation also includes several tutorials on FCIQMC, CCMC and DMQMC, which guide new users through generating the integrals (if required), running a QMC calculation along with enabling options for improving stochastic efficiency, and analysing the calculations. The HANDE source code is also heavily commented and contains extensive explanations on the theories and methods implemented (especially for CCMC), and data structures. Each procedure also begins with a comment block describing its action, inputs and outputs. We find this level of developer documentation to be extremely important for onboarding new developers and making HANDE accessible to modifications by other researchers.

\subsection{Development methodology}

The HANDE-QMC project is managed using the Git distributed version control system.\bibnote{
Git is a very powerful distributed version control system and has become the
\emph{de facto} standard in software development. Its decentralized model has
undoubtedly contributed to the huge growth of open-source software.
The official documentation is available online
\url{https://git-scm.com/}, but, as with any powerful tool, it is not an easy
task to become familiar with it. A large number of tutorials are available
online. We can recommend \url{http://gitimmersion.com/} and \url{https://coderefinery.github.io/git-intro/}.}
A public Git repository is hosted on GitHub\cite{hande-git} and is updated with new features, improvements and bug fixes.
We also use a private Git repository for more experimental development and research; this allows for new features to be iterated upon (and potentially changed or even removed) without introducing instability into the more widely available code\bibnote{We note that access to the private repository is liberally granted.}. 
We regularly update the public version, from which official releases are made, with the changes made in the private repository.
Further details of our development practices such as our development philosophy and the extensive continuous integration set up using Buildbot\cite{buildbot} are outlined in Ref.~\citenum{HANDEdev}.

\subsection{pyhande}

Interpretation and analysis of calculation output is a critical part of computational science. While we wrote scripts for performing common analyses, such as reblocking to remove the effect of autocorrelation from estimates of the standard error, we found that users would write ad-hoc, fragile scripts for extracting other useful data, which were rarely shared and contained overlapping functionality. This additional barrier also hindered curiousity-driven exploration of results. To address this, the HANDE-QMC package includes pyhande, a Python library for working with HANDE calculation outputs. pyhande extracts metadata (including version, system and calculation parameters, calculation UUID) into a Python dictionary and the QMC output into a Pandas\cite{McKinney2017-dz} \texttt{DataFrame}, which provides a powerful abstraction for further analysis. pyhande includes scripts and functions to automate common tasks, including reblocking analysis, plateau and shoulder\cite{SpencerThom_16JCP} height estimation, stochastic inefficiency estimation\cite{Vigor2016} and reweighting to reduce the bias arising from population control\cite{Umrigar1993,Vigor2015}. We have found that the development of pyhande has aided reproducibility by providing a single, robust implementation for output parsing and common analyses, and has made more complex analyses more straightforward by providing rich access to raw data in a programmable environment. Indeed, many functions included in pyhande began as exploratory analysis in a Python shell or a Jupyter notebook. The HANDE-QMC documentation also details pyhande and the tutorials include several examples of using pyhande for data analysis. pyhande makes extensive use of the Python scientific stack (NumPy\cite{Oliphant2015-pq}, SciPy\cite{scipy}, Pandas\cite{McKinney2017-dz} and Matplotlib\cite{Hunter:2007}).

\subsection{License}

HANDE-QMC is licensed under the GNU Lesser General Public
License, version 2.1.
The LGPLv2.1 is a weak copyleft license,\cite{Rosen2004-aw,St_Laurent2004-fa}
which allows the QMC implementations to be incorporated in both open- and
closed-source quantum chemistry codes while encouraging developments and
improvements to be contributed back or made available under the same
terms.\bibnote{The full legal text of the license is available from the Free
Software Foundation:
\url{https://www.gnu.org/licenses/old-licenses/lgpl-2.1.en.html}.}
pyhande is licensed under the 3-Clause BSD License,\bibnote{Legal text available
from the Open Source Initiative:
\url{https://opensource.org/licenses/BSD-3-Clause}} in keeping with many
scientific Python packages.

\section{Example results}
\label{sec:results}

In this section we present calculations to demonstrate the core functionality
included in HANDE-QMC: we consider a small molecule (nitric oxide); the uniform electron gas in the
zero-temperature ground state and at finite temperatures;
and a periodic solid, diamond, with $\pmb{k}$-point sampling. The
supplementary material includes a tutorial on running and analyzing FCIQMC on
the water molecule in cc-pVDZ basis, which is easily accessible by deterministic
methods and can be easily performed on any relatively modern laptop.

\subsection{Computational details}

All calculations in this section were run with HANDE versions earlier than version 1.3. Integrals were generated using PySCF, Psi4 and Q-Chem. Input, output and analysis scripts are available under a Creative Commons License at \url{ https://doi.org/10.17863/CAM.31933} containing specifics on which version is used for some calculations, and which SCF program is used.

\subsection{Molecules: Nitric oxide} 

Nitric oxide is an important molecule, perhaps most notably as a
signalling molecule in multiple physiological processes. Here, we
consider NO in a cc-pVDZ basis set\cite{Dunning_89JCP}, correlating all $15$ electrons. The FCI
space size is $\sim 10^{12}$, and so is somewhat beyond the reach of exact FCI
approaches. We consider initiator FCIQMC, using a walker population of $8 \times
10^6$, which is more than sufficient to achieve an accuracy of $\sim
0.1$m$E_{\textrm{h}}$. This is then compared to CCMC results for the CCSD, CCSDT
and CCSDTQ Ans\"{a}tze.
An unrestricted Hartree--Fock (UHF) molecular orbital basis is used. The computational resources to perform this study are modest compared to state-of-the-art FCIQMC simulations, never using more than about $100$ processing cores.

\begin{figure}
\includegraphics[width=\linewidth,keepaspectratio]{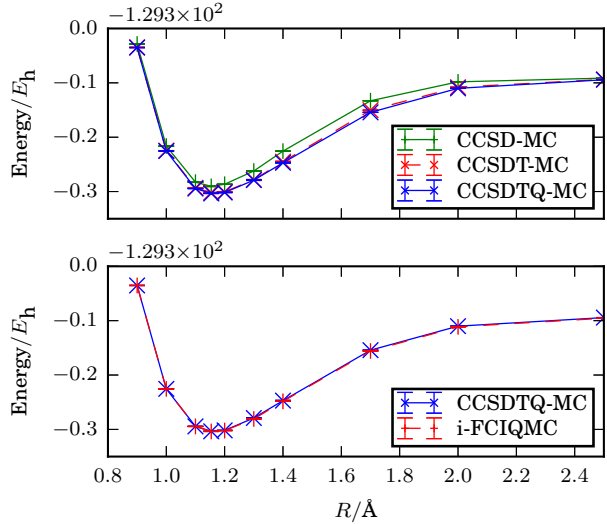}
\caption{The binding curve of NO in a cc-pVDZ basis set, correlating all electrons. Stochastic error bars are not visible on this scale, but all are smaller than $1\,\mathrm{mE_h}$. For better resolution in the differences between methods, see Table~(\ref{tab:NO}).}
\label{fig:NO}
\end{figure}

\begin{figure}
\includegraphics[width=\linewidth,keepaspectratio]{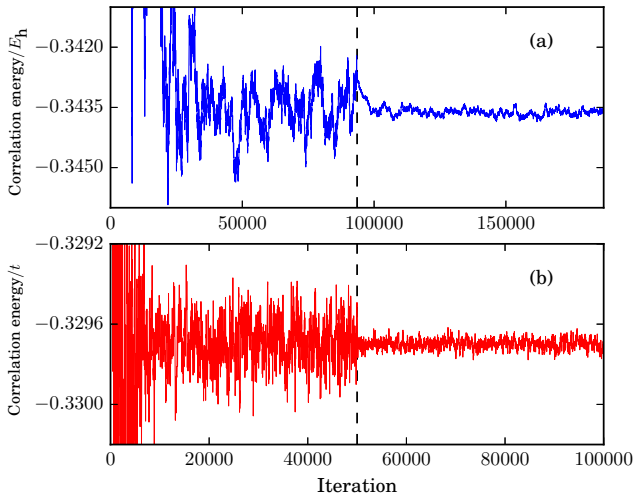}
\caption{Example simulations in HANDE-QMC using the semi-stochastic FCIQMC approach of Umrigar and co-workers\cite{Petruzielo2012}. Vertical dashed lines show the iteration where the semi-stochastic adaptation is begun, and the resulting reduction in noise is clear thereafter. (a) NO in a cc-pVDZ basis set, with all electrons correlated, at an internuclear distance of $1.154\textrm{\AA}$. The deterministic space is of size $2 \times 10^4$. (b) A half-filled two-dimensional $18$-site Hubbard model at $U/t=1.3$, using a deterministic space of size $10^4$.}
\label{fig:semistoch}
\end{figure}

In Figure \ref{fig:NO} and Table \ref{tab:NO}, results are presented for this system at varying internuclear distances. Remarkably good agreement between CCSDTQ-MC and the i-FCIQMC is achieved, with CCSDT-MC also performing extremely well. Statistical errors do not pose any issue in these results, as is typically the case for FCIQMC and CCMC simulations; all such error bars are naturally of order $0.1$m$E_{\textrm{h}}$ or less. For i-FCIQMC results the semi-stochastic adaptation was used\cite{Petruzielo2012,Blunt2015_semistoch}, choosing the deterministic space by the approach of Ref.~\citenum{Blunt2015_semistoch}. Fig.~(\ref{fig:semistoch}) demonstrates such simulations before and after enabling semi-stochastic propagation, and the benefits are clear. Indeed, i-FCIQMC results here have statistical errors of order $\sim 1 \mu E_{\textrm{h}}$ or smaller.

CCMC calculations were performed with real weights using the even selection
algorithm\cite{Scott2017}.  For the largest calculations, CCSDTQ-MC, heatbath
excitation generators were used with up to
$4.5\times10^6$ occupied excitors, parallelizing over 96 cores.  For comparison,
deterministic single reference CCSDTQ calculations performed with the MRCC
program package\cite{MRCC} required storage of $2.1\times10^7$ amplitudes, but did not converge beyond $R=1.7$\AA.  

Table~(\ref{tab:NO}) also shows the percentage of correlation energy captured by the various levels of CC, compared to i-FCIQMC. CCSD and CCSDT capture $> 92\%$ and $> 98\%$ of the correlation energy, respectively, with CCSDTQ essentially exact, and the percentage decreasing with increasing bond length as expected. The CCMC approach is particularly appropriate for such high-order CC calculations, where stochastic sampling naturally takes advantage of the sparse nature of the CC amplitudes.

\begin{table*}[t]
\begin{center}
{\footnotesize
\begin{tabular}{@{\extracolsep{4pt}}lccccccc@{}}
\hline
\hline
& \multicolumn{4}{c}{(Total energy + $129E_{\textrm{h}}$)/E$_{\textrm{h}}$} & \multicolumn{3}{c}{Correlation energy recovered ($\%$)} \\ 
\cline{2-5} \cline{6-8}
$R/\textrm{\AA}$  & CCSD  & CCSDT  & CCSDTQ  & i-FCIQMC & CCSD & CCSDT & CCSDTQ \\ 
\hline
0.9  &   -0.328507(1)  &     -0.3346(1)  &    -0.33523(4)  &   -0.335225(2)  &     97.7330(6)  &       99.78(5)  &      100.00(1)  \\
1.0  &     -0.5162(2)  &    -0.52478(2)  &   -0.525448(6)  &   -0.525470(2)  &       97.06(8)  &      99.779(6)  &      99.993(2)  \\
1.1  &   -0.582684(9)  &    -0.59317(8)  &    -0.59447(3)  &   -0.594565(3)  &      96.435(3)  &       99.58(2)  &      99.973(9)  \\
1.154  &     -0.5904(5)  &     -0.6018(3)  &     -0.6035(2)  &   -0.603772(2)  &        96.1(2)  &       99.43(9)  &       99.92(5)  \\
1.2  &    -0.58653(3)  &     -0.6005(4)  &     -0.6018(2)  &   -0.602136(3)  &      95.541(8)  &        99.5(1)  &       99.89(7)  \\
1.3  &     -0.5622(2)  &     -0.5782(4)  &     -0.5790(6)  &   -0.580833(3)  &       94.67(5)  &        99.2(1)  &        99.5(2)  \\
1.4  &     -0.5256(2)  &     -0.5451(10)  &     -0.5471(7)  &   -0.548340(3)  &       93.34(7)  &        99.1(3)  &        99.6(2)  \\
1.7  &    -0.43299(10)  &     -0.4503(5)  &     -0.4543(1)  &   -0.455765(4)  &       92.13(3)  &        98.1(2)  &       99.48(4)  \\
2.0  &    -0.39816(6)  &    -0.40800(9)  &    -0.41010(6)  &   -0.411350(2)  &       94.45(2)  &       98.59(4)  &       99.47(2)  \\
2.5  &    -0.39132(5)  &    -0.39371(8)  &    -0.39434(2)  &  -0.3954786(4)  &       98.05(2)  &       99.17(4)  &      99.467(8)  \\
\hline
\hline
\end{tabular}
}
\caption{CCMC and i-FCIQMC results for the NO molecule in a cc-pVDZ basis set, correlating all electrons, as plotted in Fig.~(\ref{fig:NO}). UHF orbitals were used. Numbers in parentheses show statistical error bars, not systematic initiator error, which is estimated to be $\sim 0.1$m$E_{\textrm{h}}$ for i-FCIQMC results. i-FCIQMC results used the semi-stochastic adaptation with a deterministic space of size $2 \times 10^4$. The results of such a semi-stochastic approach are demonstrated in Fig.~(\ref{fig:semistoch}). The final three columns show the percentage of correlation energy recovered by CCSD-MC, CCSDT-MC and CCSDTQ-MC, compared to i-FCIQMC. i-FCIQMC calculations were performed with $8 \times 10^6$ walkers, and CCMC calculations used at most $7 \times 10^6$ excips.}
\label{tab:NO}
\end{center}
\end{table*}

\subsection{Model Solid: Uniform electron gas}

HANDE also has built-in capability to perform calculations of model systems
commonly used in condensed matter physics, specifically the uniform electron gas
(UEG)\cite{Loos2016,Giuliani2005,MartinUEGChapter}, the
Hubbard model\cite{Gutzwiller1963,Hubbard1963,Kanamori1963}, and the Heisenberg model\cite{AltlandSimons, blunt_density-matrix_2014}. Such model
systems have formed the foundation of our understanding of simple solids and
strongly correlated materials, and are a useful testing ground for new
computational approaches. Studying the UEG, for example, has provided insight
into the accuracy of many-body electronic structure methods and has been a
critical ingredient for the development of many of the exchange-correlation
kernels used in Kohn--Sham density functional theory\cite{ceperley_ground_1980,perdew_self-interaction_1981,Giuliani2005dft}. 

The UEG has been used recently as a means to
benchmark and test performance of new methods, such as modifications to
diffusion Monte Carlo (DMC), as well as low orders of coupled cluster
theory\cite{Freeman1977, Bishop1978,Bishop1982, Shepherd2012c,Shepherd2013,
  Roggero2013, SpencerThom_16JCP, McClain2016,Shepherd2016a} and
FCIQMC\cite{Shepherd2012-ki, Shepherd2012-wx,
  Neufeld2017,Luo2018,Ruggeri2018,Blunt2018}.

A recent CCMC study \cite{Neufeld2017} employing coupled cluster levels up to CCSDTQ5 used HANDE to compute the total energy of the UEG at $r_s=[0.5,5]\bohr$, the range relevant to electron densities in real solids\cite{MartinUEGChapter}. The results suggest that CCSDTQ might be necessary at low densities beyond $r_s=3\bohr$\cite{Neufeld2017} in order to achieve chemical accuracy, whilst CCSDTQ5 was necessary to reproduce FCIQMC to within error bars\cite{Shepherd2012-ki, Shepherd2012-wx, Neufeld2017, Luo2018}.

HANDE was also used in the resolution of a discrepancy between restricted path-integral Monte Carlo and configuration path-integral Monte Carlo data for the exchange-correlation energy of the UEG necessary to parametrize DFT functionals at finite temperature.\citep{malone_accurate_2016,BrownUEG1,SchoofPRL,GrothUEG,DornheimUEG,DornheimPlsm,BrownUEG2,KarasievPRL,DornheimPRL,GrothPRL}. 
The UEG at finite temperatures is parametrized by the density and the degeneracy temperature, $\Theta = T/T_F$, where $T_F$ is the Fermi temperature\citep{DORNHEIM20181}. When both $r_s\approx 1 $ and $\Theta \approx 1$ the system is said to be in the warm dense regime, a state of matter which is to found in planetary interiors\cite{Fortney2009} and can be created experimentally in inertial confinement fusion experiments\cite{PhysRevB.84.224109}.

\begin{figure}
\includegraphics[width=\linewidth,keepaspectratio]{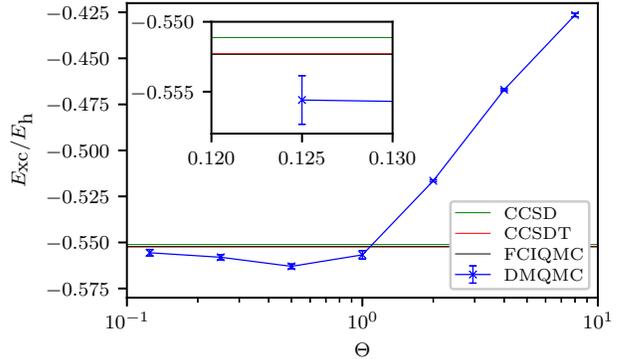}
\caption{The exchange-correlation energy ($E_{\textrm{xc}}$) for the UEG at $r_s=1\bohr$ as a function of temperature $\Theta$ using DMQMC (Ref.~\citenum{malone_accurate_2016}). The horizontal lines represent basis set extrapolated CCSD, CCSDT and FCIQMC exchange-correlation energies energies (Ref.~\citenum{Neufeld2017}). Error bars on CCMC and FCIQMC results are too small to be seen on this scale. CCSDT and FCIQMC values cannot be distinguished on this scale.
See Table.~(\ref{tab:UEG}) for numerical values for CCSD to CCSDTQ5 in the ground state.}
\label{fig:UEGrs1}
\end{figure}

Here, we show that use of HANDE can facilitate straightforward benchmarking of model systems at both zero and finite temperature.
In \cref{fig:UEGrs1} we compare DMQMC data for the 14-electron, spin-unpolarized UEG at finite $\Theta$ to zero temperature ($\Theta=0$) energies found using CCMC and FCIQMC\cite{Neufeld2017} for $r_s = 1\bohr$.
We compute the exchange-correlation internal energy
\begin{equation}
    E_{\mathrm{XC}}(\Theta) = E_{\mathrm{QMC}}(\Theta)  - T_0(\Theta),
\label{eq:xcenergy}
\end{equation}
where $E_{\mathrm{QMC}}(\Theta)$ is the QMC total energy of the UEG and $T_0$ is the ideal kinetic energy of the same UEG.
Even at $r_s=1\bohr$, coupled cluster requires contributions from triple excitations to obtain FCI-quality energies; CCSD differs by about 1mHa.
DMQMC results tend to the expected zero temperature limit given by both FCI and CC. Ground-state values from coupled cluster and FCIQMC are presented in Table.~(\ref{tab:UEG}), to make the small differences between high-accuracy methods clearer.

\begin{table}[t]
\begin{center}
{\footnotesize
\begin{tabular}{@{\extracolsep{4pt}}lc@{}}
\hline
\hline
Method  & $E_{\textrm{xc}} / E_{\textrm{h}}$ \\
\hline
CCSD-MC    & -0.551128(6) \\
CCSDT-MC   & -0.55228(1) \\
CCSDTQ-MC  & -0.55231(1) \\
CCSDTQ5-MC & -0.55232(1) \\
FCIQMC     & -0.55233(1) \\
\hline
\hline
\end{tabular}
}
\caption{Ground-state exchange-correlation energies ($E_{\textrm{xc}}$) for the UEG at $r_s=1\bohr$, comparing various levels of coupled cluster theory with FCIQMC. Exchange-correlation energies were calculated using data from Ref.\cite{Neufeld2017}.}
\label{tab:UEG}
\end{center}
\end{table}

\subsection{Solids: Diamond}
Finally, we apply HANDE-QMC to a real periodic solid, diamond, employing $\pmb{k}$ point sampling.
CCMC has been applied to 1$\times$1$\times$1 (up to CCSDTQ), 2$\times$1$\times$1
(up to CCSDT), 2$\times$2$\times$1 and 2$\times$2$\times$2 (up to CCSD) $\pmb{k}$ point meshes and non-initiator FCIQMC
to a 1$\times$1$\times$1 $\pmb{k}$ point mesh in a GTH-DZVP\cite{VandeVondele2005}\footnote[2]{As used in PySCF\cite{Sun2018}, and CP2K\cite{Hutter2014}, \url{https://www.cp2k.org/}.}
basis, and a GTH-pade pseudo-potential\cite{Goedecker1996,Hartwigsen1998}. There were 2 atoms, 8 electrons in 52
 spinorbitals per $\pmb{k}$ point. Integral files have been generated with PySCF\cite{Sun2018} using Gaussian density fitting\cite{Sun2017a}.
 Orbitals were obtained from density functional theory using the LDA
 Slater-Vosko-Vilk-Nusair (SVWN5) exchange-correlation functional\cite{Vosko1980a}
 to write out complex valued integrals at different $\pmb{k}$ points, and HANDE's read--in
  functionalities were adapted accordingly. Details of this will be the subject of a future publication on solid-state
 calculations. The \textit{heat bath uniform singles}\cite{holmes_efficient_2016,Neufeld2018} or the \textit{heat bath Power--Pitzer ref.} excitation generator\cite{Neufeld2018}
 and even selection\cite{Scott2017} or multi-spawn\cite{Spencer2018} sampling were used.
 \par Deterministic coupled cluster has been
applied to diamond previously;
Booth et al.\cite{Booth2013} have investigated diamond with CCSD, CCSD(T)\cite{Raghavachari1989} and FCIQMC
in a basis of plane waves with the projector augmented wave method\cite{Blochl1994}; McClain et al.\cite{McClain2017} studied diamond with CCSD using
GTH pseudo-potentials in DZV, DZVP, TZVP basis sets\cite{Goedecker1996,Hartwigsen1998,VandeVondele2005}\footnotemark[2];
Gruber et al.\cite{Gruber2018} used CCSD with (T) corrections in an MP2 natural orbital basis\cite{Gruneis2011}. 
\par
\begin{figure}
\includegraphics[width=\linewidth,keepaspectratio]{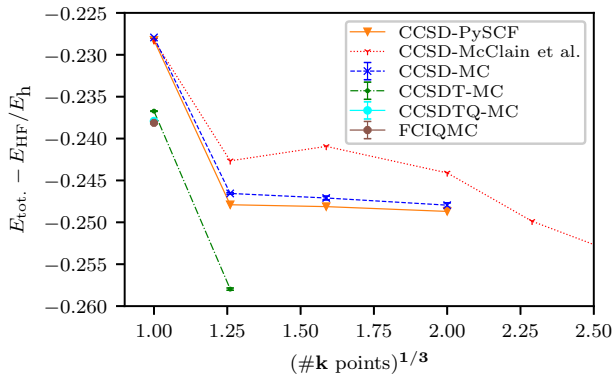}
\caption{Difference between the total and Hartree-Fock energy per $\pmb{k}$ point for diamond using CCMC (CCSD to CCSDTQ) and
	(non-initiator) FCIQMC based on
DFT orbitals. The CCSDTQ and the FCIQMC data point overlap
to a large extent. The CCSD-PySCF data was run with Hartree-Fock orbitals. In the case of CCMC, FCIQMC and CCSD-PySCF the mesh has been shifted to contain the $\Gamma$ point.
CCSD-McClain et al. is data from Figure 1 in McClain et al.\cite{McClain2017} using PySCF; we show only their data up to 12 k-points for comparison.
Both studies used the DZVP basis set and GTH pseudopotentials.}
\label{fig:diamond}
\end{figure}

 The lattice constant was fixed to 3.567\AA, as in the study by
McClain et al.\cite{McClain2017}. Figure \ref{fig:diamond} shows the
correlation energy as a function of number of $\pmb{k}$ points comparing the CCMC and FCIQMC results to the CCSD results obtained using PySCF and the CCSD results of McClain et al.\cite{McClain2017}. The correlation energy given here is calculated with respect to the HF energy, as the correlation energy from using DFT orbitals,
added to the difference of energy of reference determinant consisting of DFT orbitals and HF SCF energy.
Differences in convergences are due to the use of differently optimized orbitals, and a different treatment of the exchange integral (which will feature in a future publication). In the case of CCMC, FCIQMC and CCSD-PySCF the $\pmb{k}$ point mesh has been shifted to contain the $\Gamma$ point, while McClain et al.\cite{McClain2017} used $\Gamma$ point centered (not shifted) meshes, which explains the larger difference between CCSD-McClain et al. and the rest of the data. 
An accuracy of (0.01-0.1) eV/unit ((0.00037-0.0037)  $\mathrm{E}_h$/unit) might be required to accurately predict, for example crystal structures\cite{Wagner2016}, so these limited $\pmb{k}$-point mesh results suggest that at least CCSDT level is required for reasonable accuracy, possibly CCSDTQ. Nonetheless, we
have not considered larger basis sets, additional $\pmb{k}$ points, and other important aspects required for an exhaustive study.

\section{Discussion}
\label{sec:discussion}

This article has presented the key functionality included in HANDE-QMC: efficient, extensible implementations of the full configuration interaction quantum Monte Carlo, coupled cluster Monte Carlo and density matrix quantum Monte Carlo methods. Advances such as semi-stochastic propagation in FCIQMC\cite{Petruzielo2012,Blunt2015_semistoch} and efficient excitation generators\cite{holmes_efficient_2016,Neufeld2018} are also implemented. HANDE-QMC can be applied to model systems -- the Hubbard, Heisenberg and uniform electron gas models -- as well as molecules and solids.

We have found using a scripting language (Lua) in the input file\bibnote{We note we probably would have chosen, like Psi4, to use Python via the excellent pybind11 library had HANDE-QMC been written in C++, in part due to Python already being used extensively in scientific research.} to be extremely beneficial -- for example, in running multi-stage calculations, enabling semi-stochastic propagation after the most important states have emerged, irregular output of restart files, or for enabling additional output for debugging at a specific point in the calculation. As with (e.g.) Psi4, PySCF and HORTON, we find this approach far more flexible and powerful than a custom declarative input format used in many other scientific codes.

We are strong supporters of open-source software in scientific research and are glad that the HANDE-QMC package has been used in others research in ways we did not envisage, including in the development of Adaptive Sampling Configuration Interaction (ASCI)\cite{tubman_deterministic_2016}, understanding the inexact power iteration method\cite{lu_full_2017} and in selecting the $P$ subspace in the CC(P;Q) method\cite{Deustua2017}. We believe one reason for this is that the extensive user- and developer-level documentation makes learning and developing HANDE-QMC rather approachable. Indeed, five of the authors of this paper made their first contributions to HANDE-QMC as undergraduates with little prior experience in software development or computational science. In turn, HANDE-QMC has greatly benefited from existing quantum chemistry software, in particular integral generation from Hartree--Fock calculations in Psi4\cite{parrish_psi4_2017}, Q-Chem\cite{shao_advances_2015} and PySCF\cite{Sun2018}. We hope in future to couple HANDE-QMC to such codes to make running stochastic quantum chemistry calculations simpler and more convenient. To this end, some degree in standardization of data formats to make it simple to pass data (e.g. wavefunctions amplitudes) between codes would be extremely helpful in connecting libraries, developing new methods\cite{Deustua2017} and reproducibility.

We close by echoing the views of the Psi4 developers\cite{parrish_psi4_2017}: `the future of quantum chemistry software lies in a more modular approach in which small, independent teams develop reusable software components that can be incorporated directly into multiple quantum chemistry packages' and hope that this leads to an increased vibrancy in method development.

\begin{acknowledgement}
JSS and WMCF received support under EPSRC Research Grant EP/K038141/1 and acknowledge the stimulating research environment provided by the Thomas Young Centre under Grant No. TYC-101.
NSB acknowledges St John's College, Cambridge, for funding through a Research Fellowship, and Trinity College, Cambridge for an External Research Studentship during this work.
JE acknowledges Trinity College, Cambridge, for funding through a Summer Studentship during this work.
RSTF acknowledges CHESS for a studentship.
WH acknowledges Gonville \& Caius College, Cambridge for funding through a Research Fellowship during this work.
NSB and WH are grateful to for Undergraduate Research Opportunities Scholarships in the Centre for Doctoral Training on Theory and Simulation of Materials at Imperial College funded by EPSRC under Grant No. EP/G036888/1.
FDM was funded by an Imperial College President's scholarship and part of this work was performed under the auspices of the U.S. Department of Energy (DOE) by LLNL under Contract No. DE-AC52-07NA27344.
VAN acknowledges the EPSRC Centre for Doctoral Training in Computational Methods for Materials Science for funding under grant number EP/L015552/1 and the Cambridge Philosophical Society for a studentship.
RDR acknowledges partial support by the Research Council of Norway through its
Centres of Excellence scheme, project number 262695 and through its Mobility
Grant scheme, project number 261873.
CJCS acknowledges the Sims Fund for a studentship.
JJS is currently supported by an Old Gold Summer Fellowship from the University of Iowa. JJS also gratefully acknowledges the prior support of a Research Fellowship from the Royal Commission for the Exhibition of 1851 and a production project from the Swiss National Supercomputing Centre (CSCS) under project ID s523.
WAV acknowledges EPSRC for a PhD studentship.
AJWT acknowledges Imperial College London for a Junior Research Fellowship, the Royal Society for a University Research Fellowship (UF110161 and UF160398), Magdalene College for summer project funding for M-AF, and EPSRC for an Archer Leadership Award (project e507).
We acknowledge contributions from J.~Weston during an Undergraduate Research Opportunities Scholarships in the Centre for Doctoral Training on Theory and Simulation of Materials at Imperial College funded by EPSRC under Grant No. EP/G036888/1.
The HANDE-QMC project acknowledges a rich ecosystem of open-source projects,
without which this work would not have been possible.
\end{acknowledgement}

\appendix

\section{An introductory tutorial to HANDE-QMC}

In the following we present an introductory tutorial, demonstrating how to perform basic FCIQMC and i-FCIQMC simulations with the HANDE-QMC code. More extensive tutorials, including for CCMC and DMQMC, exist in the HANDE-QMC documentation. Here we take the water molecule at its equilibrium geometry, in a cc-pVDZ basis set\cite{Dunning_89JCP} and correlating all electrons. This is a simple example, but has a Hilbert space dimension of $\sim 5 \times 10^8$, making an exact FCI calculation non-trivial to perform.

\subsection{A basic i-FCIQMC simulation}

The input file for HANDE-QMC is a Lua script. The basic structure of such an input file is shown in Fig.~(\ref{fig:input_1}).

\begin{figure}
\begin{lstlisting}
sys = read_in {
    int_file = "INTDUMP",
}

fciqmc {
    sys = sys,
    qmc = {
        tau = 0.01,
        tau_search = true,
        rng_seed = 8,
        init_pop = 500,
        mc_cycles = 5,
        nreports = 3*10^3,
        target_population = 10^4,
        excit_gen = "heat_bath",
        initiator = true,
        real_amplitudes = true,
        spawn_cutoff = 0.1,
        state_size = -1000,
        spawned_state_size = -100,
    },
}
\end{lstlisting}
\caption{An example input file for an i-FCIQMC simulation on a molecular system. The results of such a simulation are presented in Fig.~(\ref{fig:tutorial_1}).}
\label{fig:input_1}
\end{figure}
In this the system is entirely determined by the integral file, ``INTDUMP'', which stores all of the necessary $1$- and $2$-body molecular integrals. For this tutorial, the integral file was generated through the Psi4 code\cite{parrish_psi4_2017}. Both the ``INTDUMP'' file, and the Psi4 script used to generate it, are available in additional material. As discussed in the main text, the integral file may be generated by multiple other quantum chemistry packages\cite{HORTON, Sun2018, shao_advances_2015, MOLPRO}.

In general, the system may be defined by specifying additional parameters, including the number of electrons, the spin quantum number ($M_s$), the point group symmetry label, and a CAS subspace, for example:
\begin{lstlisting}
sys = read_in {
    int_file = "INTDUMP",
    nel = 10,
    ms = 0,
    sym = 0,
    CAS = {8, 23},
}
\end{lstlisting}

\begin{figure*}[t!]
\includegraphics[width=0.7\linewidth,keepaspectratio]{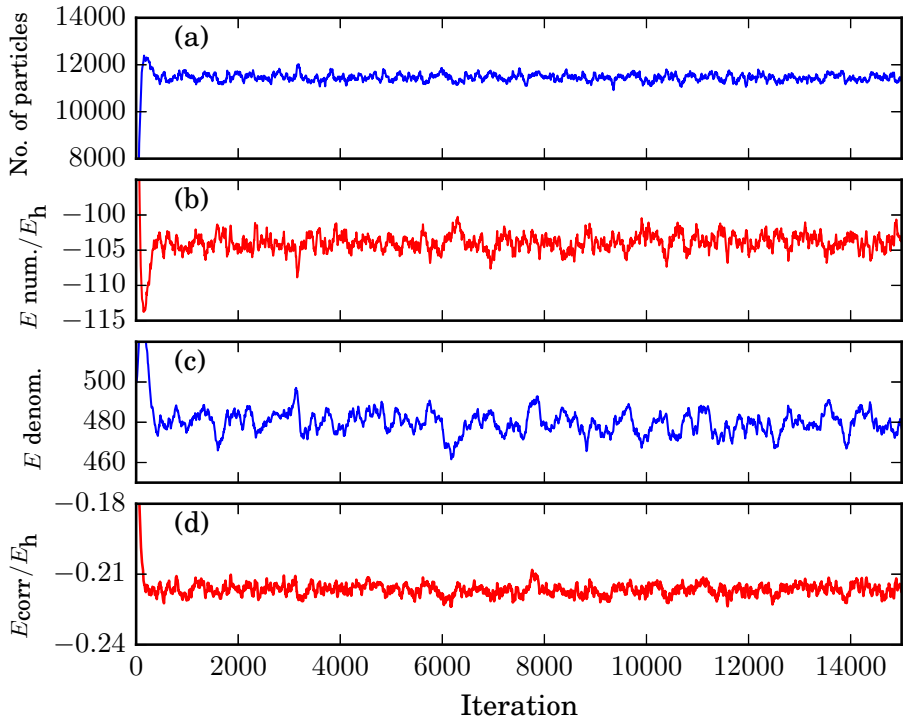}
\caption{The results of running the input file in Fig~(\ref{fig:input_1}). (a) shows the particle population, stabilizing slightly above the targeted value of $10^4$. (b) shows the numerator of the energy estimator, $\sum_{i \ne 0} C_i \braket{D_0 | \hat{H} | D_i}$, as discussed in the main text. (c) shows the energy denominator, which is the number of particles on the Hartree--Fock determinant. (d) shows the correlation energy estimates themselves.}
\label{fig:tutorial_1}
\end{figure*}

The input file then calls the \url{fciqmc{...}} function, which performs an FCIQMC simulation with the provided system and parameters. There are several options here; most are self-evident and are described in detail in the HANDE-QMC documentation. \url{tau} specifies the time step size, and \url{tau_search = true} updates this time step to an optimal value during the simulation. \url{init_pop} specifies the initial particle population, and \url{target_population} the value at which this population will attempt to stabilize. \url{excit_gen} specifies the excitation generator to be used. This option is not required, although the heat-bath algorithm of Umrigar and co-workers\cite{holmes_efficient_2016} that we have adapted for HANDE-QMC as explained in Ref.\cite{Neufeld2018}, as used here, is a sensible choice in small systems. \url{initiator = true} ensures that the initiator adaptation, i-FCIQMC, is used. \url{real_amplitudes = true} ensures that non-integer particle weights are used. This leads to improved stochastic efficiency, and so is always recommended. Lastly, \url{state_size} and \url{spawned_state_size} specify the memory allocated to the particle and spawned particle arrays, respectively - a negative sign is used to specify these values in megabytes (thus 1GB and 100MB, here).

The input file is run with
\begin{lstlisting}[frame=none, language=Bash]
$ mpiexec hande.x hande.lua > hande.out
\end{lstlisting}
with the MPI command varying between implementations in the usual way. The results of the input file in Fig.~(\ref{fig:input_1}) and presented in Fig.~(\ref{fig:tutorial_1}).

Because of the correlated nature of the QMC data, care must be taken when estimating error bars; a large number of iterations must typically be performed, allowing data to become sufficiently uncorrelated. This task can be error-prone for new users (and old ones). HANDE-QMC includes a Python script, \url{reblock_hande.py}, which performs a rigorous blocking analysis of the simulation data, automatically detecting if sufficient iterations have been performed and, if so, choosing the optimal block length to provide final estimates.

This final energy estimate can be obtained by
\begin{lstlisting}[frame=none, commentstyle=\color{black}]
$ reblock_hande.py --quiet hande.out
\end{lstlisting}

The usual estimator for the correlation energy ($E_{\textrm{corr}}$) is the Hartree--Fock projected estimator:
\begin{align}
E_{\textrm{corr}} &= \frac{ \braket{D_0 | (\hat{H} - E_{\textrm{HF}} \, \mathbb{1}) | \Psi_0§} }{ \braket{D_0 | \Psi_0} }, \\
                  &= \frac{ \sum_{i \ne 0} C_i \braket{D_0 | \hat{H} | D_i} }{ C_0 },
\end{align}
where $\ket{D_0}$ is the Hartree--Fock determinant and $E_{\textrm{HF}}$ is the Hartree--Fock energy. $C_i$ are the particle amplitudes, with $C_0$ being the Hartree--Fock amplitude. Because both the numerator and denominator are random variables, they should be averaged separately, \emph{before} performing division. It is therefore important that data be averaged from the point where both the numerator and denominator have converged individually; in some cases the energy itself may appear converged while the numerator and denominator are still converging. This does not occur in the current water molecule case, as can be seen in Fig.~(\ref{fig:tutorial_1}), where the numerator and denominator are plotted in (b) and (c), respectively. Here, all relevant estimates appear converged by iteration $\sim 1000$.

The \url{reblock_hande.py} script will automatically detect when the required quantities have converged, in order to choose the iteration from which to start averaging data. However, a starting iteration may be manually provided using \url{--start}. In general it is good practice to manually plot simulation data, as in Fig.~(\ref{fig:tutorial_1}), to check that behavior is sensible. In this case, the \url{reblock_hande.py} script automatically begins averaging from iteration number $1463$, which is appropriate.

\begin{figure}
\begin{lstlisting}
sys = read_in {
    int_file = "INTDUMP",
}

targets = {2*10^3, 4*10^3, 8*10^3, 1.6*10^4, 3.2*10^4, 6.4*10^4, 1.28*10^5}

for i,target in ipairs(targets) do
    fciqmc {
        sys = sys,
        qmc = {
            tau = 0.01,
            rng_seed = 8,
            init_pop = target/20,
            mc_cycles = 5,
            nreports = 3*10^3,
            tau_search = true,
            target_population = target,
            excit_gen = "heat_bath",
            initiator = true,
            real_amplitudes = true,
            spawn_cutoff = 0.1,
            state_size = -1000,
            spawned_state_size = -100,
        },
    }
end
\end{lstlisting}
\caption{An example input file showing how to use Lua features to perform multiple simulations in a single input file, with particle populations from $2000$ to $128,000$.}
\label{fig:input_2}
\end{figure}

\begin{table*}[t]
\begin{center}
{\footnotesize
\begin{tabular}{@{\extracolsep{4pt}}llrrrrrr@{}}
\hline                                    
\hline                                    
          &    &      Block from &   \# H psips & $\sum H_{0j} N_j$ &       $N_0$ &      Shift & Proj. Energy  \\
hande.out &  0 &  1.83000000e+03 &    2292(4)   &     -36.77(8)   &  172.6(5)   &  -0.210(3) &   -0.2131(3)  \\  
          &  1 &  1.81800000e+03 &    4602(5)   &      -56.4(1)   &  262.5(6)   &  -0.213(2) &   -0.2148(2)  \\
          &  2 &  1.47300000e+03 &    9108(7)   &     -88.29(9)   &  408.2(5)   &  -0.213(1) &   -0.2163(2)  \\
          &  3 &  1.78100000e+03 &  19050(10)   &     -151.5(1)   &  697.0(6)   &  -0.217(1) &   -0.2173(2)  \\
          &  4 &  1.97200000e+03 &  38150(10)   &     -276.7(1)   & 1270.0(6)   & -0.2188(5) &  -0.21784(6)  \\
          &  5 &  2.06500000e+03 &  74310(30)   &     -528.3(2)   &   2428(1)   & -0.2193(6) &  -0.21761(8)  \\
          &  6 &  1.82500000e+03 & 152900(30)   &    -1081.4(4)   &   4964(2)   & -0.2186(4) &  -0.21787(5)  \\

\hline                                    
\hline                                    
\end{tabular}
}
\caption{Output of the HANDE-QMC reblocking script, on the simulation with the input file of Fig.~(\ref{fig:input_2}). The final column gives the estimates of the correlation energy, as determined from the projected energy estimator.}
\label{tab:tutorial}
\end{center}
\end{table*}

\begin{figure}[t!]
\includegraphics[width=\linewidth,keepaspectratio]{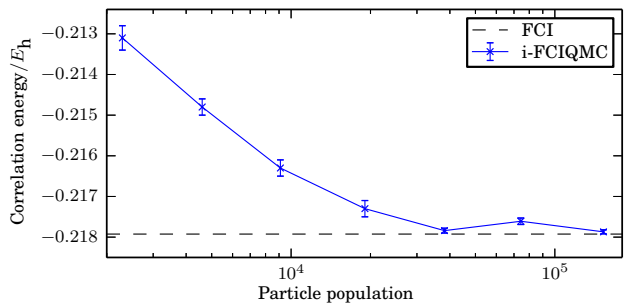}
\caption{Initiator convergence for the water molecule in a cc-pVDZ basis set, with all electrons correlated. Results were obtained by running the input file of Fig.~(\ref{fig:input_2}).}
\label{fig:tutorial_2}
\end{figure}

\subsection{Converging initiator error}

After running the \url{reblock_hande.py} script, the correlation energy estimate can be read off simply as $E_{\textrm{corr}} = -0.2166(2)E_{\textrm{h}}$. This compares well to the exact FCI energy of $E_{\textrm{FCI}} = -0.217925E_{\mathrm{h}}$, in error by $\sim 1.3\textrm{m}E_{\textrm{h}}$, despite using only $\sim 10^4$ particles to sample a space of dimension $\sim 5 \times 10^8$.

Nonetheless, an important feature of i-FCIQMC is the ability to converge to the exact result by varying only one parameter, the particle population. This is possible by running multiple i-FCIQMC simulation independently. However, one can make use of the Lua input file with HANDE-QMC to perform an arbitrary number of simulations with a single input file, as shown by example in Fig.~(\ref{fig:input_2}). Here, \url{targets} is a table containing particle populations from $2 \times 10^3$, and doubling until $1.28 \times 10^5$. We loop over all target populations and perform an FCIQMC simulation for each.

Running the \url{reblock_hande.py} script on the subsequent output file gives the results in Table~(\ref{tab:tutorial}). The final column gives the projected energy estimate of the correlation energy, and is plotted in Fig.~(\ref{fig:tutorial_2}), with comparison to the FCI energy. Accuracy within $1\textrm{m}E_{\textrm{h}}$ is reached with $N_{\textrm{w}} = 2 \times10^4$, and an accuracy of $0.1\textrm{m}E_{\textrm{h}}$ by $N_\textrm{w} = 2 \times 10^5$.

It is simple to perform a semi-stochastic i-FCIQMC simulation. To do this, as well as passing \url{sys} and \url{qmc} parameters to the \url{fciqmc} function, one should also pass a \url{semi_stoch} table. The simplest form for this table, which is almost always appropriate, is the following:
\begin{lstlisting}
    semi_stoch = {
        size = 10^4,
        start_iteration = 2*10^3,
        space = "high",
    },
\end{lstlisting}
The \url{"high"} option generates a deterministic space by choosing the most highly-weighted determinants in the FCIQMC wave function at the given iteration (which in general should be an iteration where the wave function is largely converged), $2 \times 10^3$ in this case. The total size of the deterministic space is given by the \url{size} parameter, $10^4$ in this case.

\section{Parallelization}
In this appendix, we describe two techniques that can optimize the FCIQMC parallelization, \textit{load balancing} and
\textit{non-blocking communication}. Parallelization of CCMC has been explained in Ref.~\citenum{Spencer2018} but does not yet make use of \textit{non-blocking communication}.

By and large, HANDE's FCIQMC implementation follows the standard parallel implementation of the FCIQMC algorithm, a more complete description of which can be found in Ref.~\citenum{Booth2014}. 
In short, each processor stores a sorted main list of instantaneously occupied determinants containing the determinant's bit string representation, the walker's weight as well as any simulation dependent flags.
For each iteration every walker is given the chance to spawn to another connected determinant, with newly spawned walkers being added to a second spawned walker array.
After evolution a collective \verb|MPI_AlltoAllv| is set up to communicate the spawned walker array to the appropriate processors.
The annihilation step is then carried out by merging the subsequently sorted spawned walker array with the main list.

During the simulation every walker needs to know which processor a connected determinant resides on but naturally can not store this mapping.
In order to achieve a relatively uniform distribution of determinants at a low computational cost, each walker is assigned to a processors $p$ as
\begin{equation}\label{eq:hash}
p(\ket{D_{\textbf{i}}}) = \mathrm{hash}(\ket{D_{\textbf{i}}}) \bmod  N_{\mathrm{p}},
\end{equation}
where $N_{\mathrm{p}}$ is the number of processors and hash is a hash function\citep{smhasher}.

\subsection{Load Balancing}
The workload of the algorithm is primarily determined by the number of walkers on a given processor, but the above hashing procedure distributes work to processors on a determinant basis.
For the hashing procedure to be effective we require that the average population for a random set of determinants to be roughly uniform.
Generally hashing succeeds in this regard and one finds a fairly even distribution of both walkers and determinants.
When scaling a problem of a fixed size to more processors, i.e.\ strong scaling, one observes that the distribution loses some of its uniformity with certain processors becoming significantly under and over populated which negatively affects the parallelism \citep{Booth2014}.
This is to be expected as in the limit $N_{\mathrm{p}} \rightarrow N_{\mathrm{Dets}}$ there would be quite a pronounced load imbalance unless each determinant's coefficient was of a similar magnitude (which can often be the case for strongly correlated systems).
Naturally this limit is never reached, but the observed imbalance is largely a consequence of this increased refinement.

In HANDE we optionally use dynamic load balancing to achieve better parallel performance. 
In practice, we define an array $p_{\mathrm{map}}$ as
\begin{equation}
p_{\mathrm{map}}(i) = i \bmod N_{\mathrm{p}},
\end{equation}
so that its entries cyclically contain the processor IDs, $0,\dots,N_{\mathrm{p}}-1$.
Determinants are then initially mapped to processors as
\begin{equation}\label{eq:p_map}
p(\ket{D_{\textbf{i}}}) = p_{\mathrm{map}}\Big(\mathrm{hash}(\ket{D_{\textbf{i}}})\bmod N_{\mathrm{p}} \times M\Big),
\end{equation}
where $M$ is the bin size.
\cref{eq:p_map} reduces to \cref{eq:hash} when $M = 1$.

The walker population in each of these $M$ bins on each processor can be determined and communicated to all other processors.
In this way, every processor knows the total distribution of walkers across all processors.
In redistributing the $N_{\mathrm{p}} \times M$ bins we adopt a simple heuristic approach by only selecting bins belonging to processors whose populations are either above or below a certain user defined threshold.
By redistributing bins in order of increasing population we can, in principle, isolate highly populated determinants while also allowing for a finer distribution.

This procedure translates to a simple modification of $p_{\mathrm{map}}$ so that its entries now contain the processor IDs which give the determined optimal distribution of bin.

Finally, the walkers which reside in the chosen bins have to be moved to their new processor, which can simply be achieved using a communication procedure similar to that used for the annihilation stage.
Some care needs to be taken that all determinants are on their correct processors at a given iteration so that annihilation takes place correctly.

Once the population of walkers has stabilised the distribution across processors should be roughly constant, although small fluctuations will persist.
With this in mind redistribution should only occur after this stabilisation has occurred and also should not need to occur too frequently.
This ensures that the computational cost associated with performing load balancing is fairly minor in a large calculation.
Additionally as $M$ is increased the optimal distribution of walkers should be approached, although with an increase in computational effort.

\subsection{Non-blocking communication}

HANDE also makes use of non-blocking asynchronous communication to alleviate latency issues when scaling to large processor counts\cite{gillanpeta}.
Using asynchronous communications is non-trivial in HANDE due to the annihilation stage of FCIQMC-like algorithms.
We use the following algorithm:
Consider the evolution of walkers from $\tau$ to $\tau + \Delta\tau$, then for each processor the following steps are carried out:
\begin{enumerate}
\item Initialise the non-blocking receive of walkers spawned onto the current processor from time $\tau$.
\item Evolve the main list to time $\tau+\Delta\tau$.
\item Complete the receive of walkers.
\item Evolve the received walkers to $\tau+\Delta\tau$.
\item Annihilate walkers spawned from the evolution of the two lists as well as the evolved received list with the main list on this processor.
\item Send remaining spawned walkers to their new processors.
\end{enumerate}

While this requires more work per iteration, it should result in improved efficiency if the time take to complete this work is less than the latency time.
This also ensures faster processors can continue doing work, i.e.\ evolving the main list, while waiting for other processors to finish evolving their main lists.
For communications to be truly overlapping the slowest processor would need to complete the steps above before the fastest processor reaches step (3), otherwise there will be latency as the received list cannot be evolved before all walkers spawned onto a given processor are received.

It should be pointed out that walkers spawned onto a processor at time $\tau$ are only annihilated with the main list after evolution to $\tau+\Delta\tau$, which differs from the normal algorithm.
While annihilation is vital to attaining converged results \citep{BoothAlavi_09JCP,Spencer2012} the times at which it takes place is somewhat arbitrary, once walkers are annihilated at the same point in simulation time.
Communication between processors is also required when collecting statistics, however the usual collectives required for this can simply be replaced by the corresponding non-blocking procedures.
This does require that information is printed out in a staggered fashion but this is of minor concern.

\bibliography{hande}

\begin{mcitethebibliography}{149}
\bibitem[Not()]{Note-1}
The original algorithm used integer weights. It was subsequently shown that
  floating-point weights greatly reduce the stochastic noise. HANDE-QMC uses
  fixed-precision for the weights such that both approaches can be
  straightforwardly handled.\relax
\mciteBstWouldAddEndPunctfalse
\mciteSetBstMidEndSepPunct{\mcitedefaultmidpunct}
{}{\mcitedefaultseppunct}\relax
\EndOfBibitem
\bibitem[Not()]{Note-3}
DMQMC averages over independent calculations and so does not suffer from a
  correlation issue.\relax
\mciteBstWouldAddEndPunctfalse
\mciteSetBstMidEndSepPunct{\mcitedefaultmidpunct}
{}{\mcitedefaultseppunct}\relax
\EndOfBibitem
\bibitem[Not()]{Note-5}
We note that access to the private repository is liberally granted.\relax
\mciteBstWouldAddEndPunctfalse
\mciteSetBstMidEndSepPunct{\mcitedefaultmidpunct}
{}{\mcitedefaultseppunct}\relax
\EndOfBibitem
\bibitem[Not()]{Note-8}
We note we probably would have chosen, like Psi4, to use Python via the
  excellent pybind11 library had HANDE-QMC been written in C++, in part due to
  Python already being used extensively in scientific research.\relax
\mciteBstWouldAddEndPunctfalse
\mciteSetBstMidEndSepPunct{\mcitedefaultmidpunct}
{}{\mcitedefaultseppunct}\relax
\EndOfBibitem
\end{mcitethebibliography}

\end{document}